\documentclass[12pt]{article}

\usepackage[latin1]{inputenc}
\usepackage{amsmath}
\usepackage{amsfonts}
\def\be{\begin{equation}}
\def\ee{\end{equation}}
\def\ba{\begin{eqnarray}}
\def\ea{\end{eqnarray}}

\begin{document}


\newpage
\thispagestyle{empty}
\begin{center}
\vspace{0.4cm}
{\Large {\bf STRICTLY CANONICAL QUANTIZATION }}\\
\vspace{0.3cm}
{\Large {\bf OF A MASSLESS SPINNING PARTICLE }}\\
\vspace{0.3cm}
{\Large {\bf AND A QUATERNIONIC EXTENSION OF }}\\
\vspace{0.3cm}
{\Large {\bf PSEUDOCLASSICAL MECHANICS }} \\
\vspace{1.0cm}
{\large {\bf Marlon Nunes Barreto}}\footnote{e-mail: marlonnunes@yahoo.com.br} \\
\vspace{0.2cm}
{\it Universidade Federal de Sergipe and Universidade Federal da Paraíba}\\
\vspace{0.5cm}
{\large {\bf Flávio Jamil Souza Ferreira\footnote{e-mail: jamilbanger@hotmail.com} and Stoian Ivanov Zlatev\footnote{e-mail: zlatev@fisica.ufs.br}}} \\
\vspace{0.2cm}
{\it Universidade Federal de Sergipe} \\
\vspace{1.0cm}
\end{center}

\begin{abstract}
A pseudoclassical model, reproducing, upon quantization, the dynamics of the chiral sectors of the massless spin-1/2 field theory is proposed. The discrete symmetries of the action are studied in details. In order to reproduce the positive and negative chiral sectors of the particle and antiparticle, we promote the algebra of functions on the phase space to a bimodule over the complexified quaternions - biquaternions. The quantization is performed by means of strictly canonical methods (Dirac brackets formalism) reproducing the Dirac equation in the Foldy-Wouthuysen representation in the particle and antiparticle sectors and the Weyl equation in the chiral sectors.
\end{abstract}

\vspace{3cm}

\newpage
\section{Introduction}

Canonical quantization is the art of establishing a correspondence - following Dirac rules - between a classical (or pseudoclassical) model and a quantum system. Its universal status is questionable, since one cannot discard the possibility quantum mechanics to exist - in some particular cases - on its own right, without any relation to some (pseudo-)classical mechanics. On the other hand, canonical quantization has proven its efficiency on a vast variety of examples. 

The inverse problem - that of finding a (pseudo-)classical system reproducing, upon quantization, given quantum dynamics - is also interesting. In particular, after this problem being solved, a path integral for the quantum system can be obtained. 

The pseudoclassical relativistic particle models \cite{bm75,casalbuoni76} generally reproduce 
the  consistent quantum mechanics \cite{gg00} of the one-particle and one-antiparticle 
sectors of the corresponding quantum field theory. 
Two cases have resisted the attempts to obtain a pseudoclassical  model yelding a consistent quantum mechanics upon  canonical quantization. These are the massive spin-1/2 particle in odd-dimensional space-time (with a given value of spin) and the  massless particle. In both cases the problem is to obtain a ``minimal'' model. Indeed, a theory, reproducing, for example, the quantum mechanics of a spin-1/2 particle with both signs of spin in (2+1)-dimensional space-time, can be easily obtained by reduction from the spinning particle model in (3+1) dimensions.  
A common technical problem arises in the both cases: a first-class bifermionic constraint which does not admit gauge fixing \cite{fggm04}.   

In the present paper we propose a pseudoclassical model for the massless spinning particles which is free of the problem. The pseudoclassical action is extracted from a path integral in a way similar to that of Refs.~\cite{fg91, gitman97}. In order to obtain a canonically quantizable model we had to solve the bifermionic constraint problem, as well as the bifermionic constant problem \cite{fggm04}. We looked for a model satifying certain condititions. The most important among them was to obtain a bifermionic constraint belonging to a set of second-class constraints, instead of being a first-class constraint. Also, we looked for a model with the simplest possible constraint structure and, in particular, leading to the same set of first-class constraints as in the (massless) spinless case.

The canonical quantization of the model reproduces the quantum mechanics of the one-fermion state and that of the negative-energy states, corresponding to the antifermion states.  

In order to reproduce the dynamics in the chiral sectors, i.e., the dynamics of states of given chirality, we propose a quaternionic generalization of the pseudoclassical mechanics. The `algebra of functions on the the phase space' is assumed to be an algebra over biquaternions (a quaternionic bimodule). This allows us to develop a pseudoclassical scheme which reproduces, upon canonical quantization, the quantum mechanics in the chiral sectors. 
The paper is organized as follows. In Sect.~2 pseudoclassical action is derived from a a path integral for the propagator. The discrete symmetries of the action are studied in Sect.~3.  The Hamiltonian  formulation is given in Sect.~4. Canonical quantization is performed in Sect.~5.

\section{The path integral}
The propagator $S^c$ of the free massless Dirac field satisfies the equation
\begin{equation}
\label{propagator1}
i\gamma^\mu\partial_\mu S^c(x)=-\delta^4(x)
\end{equation}
and Feynman assymptotic conditions. 
In order to get an action, leading to a suitable set of  constraints,
we obtain a path-integral expression for the function
\begin{equation}
\label{Delta}
\Delta(x)=\gamma^5\gamma^0 S^c(x).
\end{equation}
It can be expressed, following Schwinger \cite{schwinger51}, as a matrix element of an operator in a Hilbert space. Let $\hat{X}^\mu$, $\hat{P}_\mu$ be a set of operators, generating an irreducible representation of the canonical commutation relations
\[
\left[\hat{X}^\mu, \hat{X}^\nu\right]_-=\left[\hat{P}_\mu, \hat{P}_\nu\right]_=0, \quad
\left[\hat{X}^\mu, \hat{P}_\nu\right]_-
=i\delta^\mu_\nu.
\] 
in a Hilbert space, in such a way that 
\[
\langle x\mid \hat{P}_\mu \mid\psi\rangle= -i\partial_\mu
\langle x\mid\psi\rangle
\]
where $\mid x\rangle$ is an eigenvector for all $X^\mu$ and $\psi$ is an arbitrary vector from a suitable dense set in the Hilbert space. 
Then 
\[
\Delta(x_{\textrm{out}}-x_{\textrm{in}})=\langle x_{\textrm{out}}|\hat{\Delta}|x_{\textrm{in}}\rangle
\]
where 
\begin{equation}
\label{deltaop}
\hat{\Delta}=\frac{\gamma^5 \hat {P}^0+\gamma^5{\alpha}^i\hat{P}_{i}}
{\hat{P}^2+i\epsilon},
\end{equation}
$\alpha^k=\gamma^0\gamma^k$ ($k=1,2,3$),
and the weak limit $\epsilon\to 0+$ is understood. We omit the term $i\epsilon$ in the sequel. One can use the Schwinger proper-time representation \cite{schwinger51} for the inverse operator $1/\hat{P}^2$, 
\begin{equation}
\label{denominator}
\frac 1 {\hat{P}^2}=\frac{1}{i} \int_0^\infty d\lambda\, e^{i\lambda \hat{P}^2}.
\end{equation}
As to the factor $\gamma^5 \hat{P}^0+\gamma^5\boldsymbol{\alpha}\cdot\hat{\mathbf{P}}$, it could be 
represented by an integral over a pair of Grassmann variables \cite{gitman97}, $\chi_{1}$ and $\chi_{2}$ , 
\begin{equation}
\label{numerator1}
\gamma^5 \hat{P}^0-\gamma^5{\alpha}^{i}\hat{P}_{i}
=i\int d\chi_1d\chi_2\, \exp\left[i\chi_1\chi_2\left(\gamma^5 \hat{P}^0+\gamma^5{\alpha}^{i}\hat{P}_{i}\right)\right]. 
\end{equation}
Some other possibilities also exist. Let $B$ be some quantity, then  
\begin{equation}
\label{numerator2}
B=-\frac{1}{2\pi}\oint_{\mathcal{C}} \frac{dz}{z^2}\, 
e^{izB},
\end{equation}
where $\mathcal{C}$ is a curve circling around the origin in the $z$-plane.
Using eqs.~(\ref{denominator}), (\ref{numerator2}) and 
\[
\epsilon_{ijk}\alpha^{j}\alpha^{k}=2i\gamma^{5}\alpha^{i}
\]
in (\ref{deltaop}), one obtains
\begin{eqnarray}
\label{hatdelta1}
{\Delta}(x_{\mathrm{out}}-x_{\mathrm{in}})
&=&\frac{i}{2\pi}\int_0^\infty d\lambda \oint_{\mathcal{C}} \frac{dz}{z^2}
\langle x_{\mathrm{out}}\mid
\exp\Big[i\lambda \hat{P^{2}}\\
&&+iz\big(\gamma^5 \hat{P}^0+\frac{i}{2}\epsilon_{ijk}\hat{P}^{i}\alpha^{j}\alpha^{k}\big)\Big]\mid x_{\mathrm{in}}\rangle.
\nonumber
\end{eqnarray}
The integrand can be considered as an evolution operator for   a quantum-mechanical system with Hamiltonian
\begin{equation}
\label{hamiltonian}
\mathcal{H}(\hat P, \lambda, z)= 
-s \hat{P^{2}}-z\left(\gamma^5 \hat{P}^0+\frac{i}{2}\epsilon_{ijk}\hat{P}^{i}\alpha^{j}\alpha^{k}\right). 
\end{equation}
A path integral  for the matrix element in (\ref{hatdelta1})
is obtained in the standard way \cite{fg91,gitman97}. 
Using the completness relation 
\[
\int d^4x |x\rangle\langle x|=I, 
\]
and introducing $\delta$-functions in $s$, $u$ and $v$, one obtains
\begin{eqnarray}
\label{bosonic_integral1}
&&\langle x_{\mathrm{out}}\mid\exp\big[-i\mathcal{H}(\hat{P},s,z)\big]\mid x_{\mathrm{in}}\rangle
=
\int d^4x_{1}\dots \int d^4x_{N-1}
\nonumber \\
&&\times\int_0^\infty ds_{1}\dots \int_0^\infty ds_{N}
\,\delta(s_{N}-s_{N-1})\dots 
\delta(s_{1}-s_{0})
\nonumber\\
&&\times\int du_{1}\dots \int du_{N}
\,\delta(u_{N}-u_{N-1})\dots 
\delta(u_{1}-u_{0})\\
&&\times \int dv_{1}\dots \int dv_{N}
\,\delta(v_{N}-v_{N-1})\dots 
\delta(v_{1}-v_{0})
\nonumber \\
&&\langle x_{\textrm{out}}\mid \exp\{-i\mathcal{H}(\hat{P},s_{N}, u_{N}+iv_{N})\Delta_{N} \tau\}\mid x_{N-1}\rangle \dots 
\nonumber\\
&&\langle x_{1}\mid\exp\{-i\mathcal{H}(\hat{P},s_{1}, u_{1}+iv_{1})\Delta_1\tau\}\mid x_{\mathrm{in}}\rangle,
\nonumber
\end{eqnarray}
where 
$\sum_{i=1}^N\Delta_i\tau=1$. Evaluating the matrix element, 
\begin{eqnarray*}
&&\langle x_{k}\mid\exp\Big[-i\mathcal{H}(\hat{P},
s_{k}, u_{k}+iv_{k})\Delta_k\tau\Big]\mid x_{k-1}\rangle\\
&&=\int\frac{d^4p_{k}}{(2\pi)^4}
\exp\Big[ip_{k}(x_{k}-x_{k-1})-i\mathcal{H}(\hat{P},s_{k}, u_{k}+iv_{k})\Delta_k\tau\Big]
\end{eqnarray*}
and using integral representations for the $\delta$-functions
one obtains
\begin{eqnarray}
\label{bosonic_integral2}
&&\langle x_{\mathrm{out}}\mid\exp\big[-i\mathcal{H}(\lambda,z)\big]\mid x_{\mathrm{in}}\rangle
\nonumber \\
&&=
\int d^4x_{1}\dots \int d^4x_{N-1}
\int \frac{d^4p_{1}}{(2\pi)^4}\dots
\int\frac{d^4p_{N}}{(2\pi)^4}
\nonumber \\
&&\times
\int_0^{+\infty} ds_{1}\dots \int_{0}^{+\infty} ds_{N}
\int_{-\infty}^{+\infty} \frac {dP_{s1 }}{2\pi}\dots 
\int_{-\infty}^{+\infty} \frac {dP_{sN}}{2\pi}
\nonumber \\
&&\times\int_{-\infty}^{+\infty} dh_{1}\dots \int_{-\infty}^{+\infty} dh_{N}
\int_{-\infty}^{+\infty} {dP_{h1}}\dots 
\int_{-\infty}^{+\infty} {dP_{hN}}
\nonumber \\
&&\times\int_{-\infty}^{+\infty} du_{1}\dots \int_{-\infty}^{+\infty} du_{N}
\int_{-\infty}^{+\infty} {dP_{u1}}\dots 
\int_{-\infty}^{+\infty} {dP_{uN}}
 \\
&&\times\int_{-\infty}^{+\infty} dv_{1}\dots \int_{-\infty}^{+\infty} dv_{N}
\int_{-\infty}^{+\infty} {dP_{v1}}\dots 
\int_{-\infty}^{+\infty} {dP_{vN}}
\nonumber \\
&&\times \mathrm{T} \exp\Big\{i\sum_{k=1}^N\Big[
p_{k}\frac{x_{k}-x_{k-1}}{\Delta_{k}\tau}+
{P_{sk}}\frac{s_{k}-s_{k-1}}{\Delta_k\tau}
+P_{hk}\frac{h_{k}-h_{k-1}}{\Delta_k\tau}
\nonumber \\
&&+P_{uk}\frac{u_{k}-u_{k-1}}{\Delta_k\tau}
+P_{vk}\frac{v_{k}-v_{k-1}}{\Delta_k\tau}
-\mathcal{H}(p_{k},  
s_{k}, u_{k}+iv_{k}) 
\Big]\Delta_k\tau \Big\},
\nonumber
\end{eqnarray}
where $u_{0}+iv_{0}=z$, $s_0=\lambda$, $h_{0}=\gamma^5$ and 
$\mathrm{T}$ indicates an ordered exponent  ($\alpha$-matrices in the factors with greater value of $k$ stay to the left of the ones with smaller $k$).
Making the rescaling $\lambda=e_{0}/2$, $s_{k}=e_{k}/2$, $P_{sk}=2P_{ek}$ and taking the limit $\Delta_k\tau\to 0$, one obtains the path integral 
\begin{eqnarray}
\label{integral1}
&&\Delta^{c}(x_{\mathrm{out}}-x_{\mathrm{in}})=\frac{i}{4\pi}\int_0^\infty d e_{0} \oint_{\mathcal{C}} \frac{dz}{z^2} 
\int_{x_{\textrm{in}}}^{x_{\textrm{out}}} Dx\int Dp 
\nonumber \\
&&\times\int_{e_{0}}De  \int DP_e \int_{h_{0}}Dh  \int DP_h   
\int_{u_{0}}Du  \int DP_u 
\int_{v_{0}}Dv  \int DP_v 
\nonumber\\
&&\times \mathrm{T}
\exp\Big\{i\int_{0}^{1}\Big[p_\mu \dot x^\mu+P_e \dot e
+P_h\dot h+P_u\dot u+P_v\dot v  \\
&&+\frac{1}{2}e p^{2}+(u+iv)\big(h p^{0} + \frac{i}{2}\epsilon_{ijk}p^{i}\alpha^{j}\alpha^{k}\big)
\Big]d\tau\Big\}\Bigg\vert_{h_{0}=\gamma^5}.
\nonumber
\end{eqnarray}
The time-ordered exponent in eq.~(\ref{integral1}) can be expressed as a linear combination of antisymmetrized products of $\alpha$-matrices by means of a path-integral technique \cite{fg91,gitman97,gzb98}) with the result
\begin{eqnarray}
\label{integral2}
&&\Delta^{c}(x_{\mathrm{out}}-x_\mathrm{in})=\frac{i}{4\pi}
\textrm{Sym}\int_0^\infty d e_{0} \oint_{\mathcal{C}} \frac{dz}{z^2}\int_{x_{in}}^{x_{out}} Dx\int Dp\int_{e_{0}}De  \int DP_e
\nonumber \\ 
&&\times\int_{h_0} Dh \int D P_h \int_{u_{0}} Du\int DP_u 
\int_{v_{0}} Dv\int DP_v \int_{\boldsymbol{\psi}(0)+\boldsymbol{\psi}(1)=\boldsymbol{\theta}} 
\mathcal{D}\boldsymbol{\psi}
\nonumber
\\
&&\times
\exp\Bigg\{i\int_{0}^{1}\Bigg[ 
P_e \dot{e}+P_u\dot{u}+P_v\dot{v}+P_h\dot{h}+p_{\mu}\dot x^{\mu}
+i\boldsymbol{\psi}\cdot\dot{\boldsymbol{\psi}}
+\\
&&+\frac{1}{2}e p^{2}+(u+iv)\left(h p_{0} + 2i\epsilon_{ijk}p^{i}\psi^{j}\psi^{k}\right)
\Bigg]d\tau-\boldsymbol{\psi}(1)\cdot \boldsymbol{\psi}(0)\Bigg\}\Bigg|_{
\boldsymbol{\theta}=\boldsymbol{\alpha}, \, h_0=\gamma^{5}},
\nonumber
\end{eqnarray}
where $\boldsymbol{\psi}(\tau)$ is a Grassmann-odd trajectory, $\theta^k$ are Grassmann-odd parameters, and the translationally-invariant measure $\mathcal{D}\boldsymbol{\psi}$ is normalized, 
\[
\int_{\boldsymbol{\psi}(0)+\boldsymbol{\psi}(1)=0}
\mathcal{D}\boldsymbol{\psi}\,\exp\left(
-\int_{0}^{1}\boldsymbol{\psi}\cdot\dot{\boldsymbol{\psi}}\,d\tau\right)=1.
\]
Performing the integration over the trajectories in the momentum space $p$ one obtains the path integral in the Lagrangian form 
\begin{eqnarray} 
\label{integral-l}
&&\Delta^{c}(x_{out},x_{in})=\frac{i}{4\pi}\textrm{Sym}\int_0^\infty d e_{0} \oint_{\mathcal{C}} \frac{dz}{z^2}\int_{x_{in}}^{x_{out}} Dx\, M(e) \int_{\boldsymbol{\psi}(0)+\boldsymbol{\psi}(1)=\boldsymbol{\theta}}\mathcal{D}\boldsymbol{\psi}
\nonumber \\
&&\times\int_{e_{0}}De  \int DP_e \int_{h_0} Dh \int D P_h  
\int_{u_{0}} Du\int DP_u\int_{v_{0}} Dv\int DP_v  
\nonumber
\\
&&\times
\exp\Bigg\{i\int_{0}^{1}\Bigg[P_e \dot e+P_u\dot u+P_v\dot v +P_h\dot h+
p_\mu \dot x^\mu
\nonumber 
\\
&&+i\boldsymbol{\psi}\cdot\dot{\boldsymbol{\psi}}
-\frac{1}{2e} \tilde{w}^{2}\Bigg]d\tau-\boldsymbol{\psi}(1)\cdot {\boldsymbol{\psi}}(0)\Bigg\}\Bigg|_{\boldsymbol{\theta}=\boldsymbol{\alpha}, h_0=\gamma^5},
\end{eqnarray}
where
\begin{equation}
\tilde{w}^{0}=\dot{x}^{0}+ (u+iv)h, \qquad
\tilde{w}^k=\dot{x}^k+2i(u+iv)\epsilon_{klm}\psi^l\psi^m,
\end{equation}
and
\begin{equation}
M(e)=\int Dp\,\exp\left\{ \frac{i}{2}\int_{0}^{1}ep^{2}d\tau \right\}.
\end{equation}

\newpage

\section{The action and its symmetries}
The gauge-fixing terms $P_{e}\dot{e}$, $P_{u}\dot{v}$, $P_{v}\dot{v}$, $P_{h}\dot{h}$ and the symbol-defining term $-\boldsymbol{\psi}(1)\cdot {\boldsymbol{\psi}}(0)$ apart, the argument of the exponent in (\ref{integral-l}) is the action. Since the complex nature of the variable $z=u+iv$ is not essential, we put $v=0$ and the action takes the form  
\begin{equation}
\label{action}
S=\int_{0}^{1}Ld\tau=\int_{0}^{1}\left(-\frac{w^{2}}{2e}+i\boldsymbol{\psi}\cdot\dot{\boldsymbol{\psi}}\right)d\tau,
\end{equation}
where
\begin{equation}
{w}^{0}=\dot{x}^{0}+ uh, \qquad
{w}^k=\dot{x}^k+2iu\epsilon_{klm}\psi^l\psi^m.
\end{equation}

The action (\ref{action}) is reparametrization invariant. It is also invariant under space reflection $\mathcal{P}$ if the variables are  transformed as follows
\begin{equation}
\label{space}
\mathcal{P}:\quad x^0\to x^0, \quad x^k\to -x^k,\quad \psi^k\to -\psi^k, \quad h\to -h,\quad u\to -u,
\end{equation}
and under the time reflection $\mathcal{T}^{\prime}$,
\begin{equation}
\label{space2}
\mathcal{T}^{\,'}:\quad x^0\to -x^0, \quad x^k\to x^k,\quad \psi^k\to -\psi^k, \quad e\to e \quad h\to -h,\quad u\to u.
\end{equation}
The dynamics is invariant under the trajectory-reversal operation $\mathcal{R}$, i.e.,  
\[ 
\int^{0}_{-1} \tilde {L} d\tau=S
\]
where $\tilde{L}$ is obtained by the following substitutions in $L$, eq.~(\ref{action}): 
\begin{eqnarray}
\label{time}
\mathcal{R}:&&x^{\mu}(\tau)\to \tilde {x}^{\prime\mu}(\tau)= x^{\mu}(-\tau), \quad \psi^k(\tau)\to \tilde\psi^{k}(\tau)=-\psi^k(-\tau),
\nonumber\\
&&h(\tau)\to \tilde{h}(\tau)= -h(-\tau),\quad 
u(\tau)\to \tilde{u}(\tau)=u(-\tau), \\
&&e(\tau)\to \tilde{e}(\tau)=e(-\tau)\qquad i\to -i.
\nonumber
\end{eqnarray}

Consequently, the equations of motion are invariant under the (pseudoclassical counterpart) of the Wigner time inversion operation 
\[
\mathcal{T}=\mathcal{T}^{\prime}\mathcal{R}=\mathcal{R}\mathcal{T}^{\prime}.
\]

\section{Pseudoclassical mechanics}
The pseudoclassical mechanics can be formulated in terms of the $\mathbb{Z}_2$-graded associative algebra $\mathcal{A}$, generated by the canonical variables. For some of the non-nilpotent variables we assume the existence of the corresponding inverses. The phase space $\Gamma$ is a supermanifold \cite{DeWitt92}. We are not going to specify the structure here. However, we refer to the elements of $\mathcal{A}$ as to functions on the phase space. 

The evolution of a function on the phase space with the evolution parameter $\tau$ is governed by the Hamiltonian equations. The evolution parameter $\tau$ can also be understood as a Grassmann-even generating element in a larger algebra $\mathcal{A}^{\prime}$, generated by the canonical variables and $\tau$. The element $\tau$ has vanishing Poisson brackets with any function on the phase space. 

The constraints are understood as conditions, defining an equivalence  relation in $\mathcal{A}$. We limit ourselves to the case of $\tau$-independent constraints. Two elements $f$ and $g$ of $\mathcal{A}^{\prime}$ are considered equivalent if the difference $f-g$ is ``proportional to the constraints'', i.e., if 
\[
f-g=\sum_{a, i} F_{i}\Phi_{a}G_{i},
\]
where $\Phi_a$ ($a=1\dots N$) are the constraint functions and $F_{i}$, $G_i$ are algebra elements.  
Obviously, this define a quotient algebra $\tilde{\mathcal{A}}$ (the algebra of functions on the constraint suface).
Suppose that the set of all constraints is a second-class one. If Dirac brackets with respect of this set are used, the Hamilton equations define a consistent evolution in the quotient algebra $\tilde{\mathcal{A}}$.  

\subsection{Hamiltonian formulation}

The canonical momenta $P_e$, $P_u$, and $P_h$, conjugate to $e$, $u$, 
and $h$, correspondingly, vanish. The momenta 
conjugate to $\psi^k$ are given by
\[
\Pi_k:=\frac{\partial_r L}{\partial \dot{\psi}^k}=i\psi^k.
\]   
There are six primary constraints in the theory. The constraint functions are
\begin{equation}
\label{primary}
\Phi_{1}^{(1)}=P_{e}, \quad \Phi_{2}^{(1)}=P_u, \quad \Phi_{3}^{(1)}=P_{h},\quad 
\Phi_{4i}^{(1)}=\Pi_{i}-i\psi^{i}.
\end{equation}
and the Hamiltonian $H^{(1)}$ is given by
\begin{equation}
\label{hamiltonian1}
H^{(1)}=-\frac{1}{2} e p^{2}-u\left(h p_0+2i\epsilon_{klm}p^k\psi^{l}\psi^{m}\right)+\lambda^{\alpha}\Phi_{\alpha}^{(1)},
\end{equation}
where $\lambda^{\alpha}$ are Lagrange multipliers associated with the constraints. The constraints $\Phi_{4i}$ can be eliminated. Under a change of the fermionic variables, 
\begin{equation}
\psi^{k}=\frac 1 2 \left(\xi^k+\eta^{k}\right), \qquad 
\Pi_{k}=\frac 1 {2i} \left(\xi^{k}-\eta^{k}\right)
\end{equation}
one finds that the new variables have the following Poisson brackets:
\begin{equation}
\left\{\xi^{i},\xi^{j}\right\}=2i\delta_{ij}, \quad
\left\{\eta_{i},\eta_{j}\right\}=-2i\delta_{ij},\quad
\left\{\xi_{i},\eta_{j}\right\}=0.
\end{equation}
and $\Phi_{4i}^{(1)}=\xi^i$.
The conservation of the constraint functions $\Phi_{4i}^{(1)}$ implies $\lambda^{4i}=0$ and the terms in $H^{(1)}$, linear in $\xi^k$, vanish. On the other hand, the Poissonn brackets with all the variables of the terms in $H^{(1)}$, bilinear in $\xi^k$,  vanish on the constraint surface and can be omitted. The Hamiltonian takes the form
\begin{equation}
\label{hamiltonian2}
H^{(2)}=-\frac{1}{2} ep^{2}-u\left( hp_{0}+\frac{i}{2}
\epsilon_{ijk}p^{i}\eta^{j}\eta^{k}\right)+\sum_{\alpha=1}^{3}
\lambda^{\alpha}\Phi_{\alpha}^{(1)}.
\end{equation}
The conservation of the primary constraints $\Phi^{(1)}_1$, $\Phi^{(1)}_2$, and $\Phi^{(1)}_3$ yield three secondary constraints, 
\begin{eqnarray}
\label{secondary}
\Phi_{1}^{(2)}= p^{2},\qquad  
\Phi_{2}^{(2)}=  hp_{0}+\frac{i}{2}\epsilon_{ijk}p^{i}
\eta^{j}\eta^{k}\qquad \Phi_{3}^{(2)}=u.
\end{eqnarray}
The consistency conditions on the constraint functions $\Phi_{2}^{(2)}$, $\Phi_{3}^{(2)}$ imply 
$\lambda^3=\lambda^2=0$.
The only Lagrange multiplier remaining unfixed is $\lambda^1$.

One can separate the constraints into a set of first-class constraints, 
\begin{equation}
\label{firstclass}
\Phi_{1}^{(1)}=P_{e}, \qquad \Phi_{1}^{(2)}= p^2,
\end{equation}
and a second-class set of constraints, 
\begin{eqnarray}
\label{secondclass}
\Phi_{2}^{(1)}=P_{u}, &&
\qquad \Phi_{3}^{(2)}=u,\\
\Phi_{3}^{(1)}=P_{h}, &&
\Phi_{2}^{(2)}= hp_{0}+\frac{i}{2}\epsilon_{ijk}p^{i}
\eta^{j}\eta^{k}.
\end{eqnarray}
The set of the first class constraints is exactly the one appearing in the theory of scalar massless particles. 
Only one constraint function depends on the Grassmann-odd variables $\eta^k$ and it belong to the set of second-class constraints. 

In this stage $\Phi^{(2)}_{2}$ is already contained in a set of second-class constraints, that is, without using any gauge condition we get to solve the bifermionic constraint problem \cite{fggm04}.

\subsection{Gauge fixing}
With the aim to use the physical time $x^0$ as evolution parameter, we impose the gauge 
condition
\begin{equation}
\Phi_{1}^{G}=0, \qquad \textrm{where}\qquad \Phi_{1}^{G}=x^{0}-\tau.
\end{equation}
The constraint function $\Phi_{1}^{G}$ depends on $\tau$. In order to obtain a $\tau$-indepent set of constraints we make a  canonical transformation \cite{gt90}. The generating function is 
\[
W=x^{\mu}p^{\prime}_{\mu}-\tau p^{\prime}_0,
\]
so that the relation between the new variables $x^{\prime \mu}$,
$p^{\prime}_{\mu}$ and the old ones, $x^{\prime \mu}$,
$p^{\prime}_{\mu}$, is given by
\begin{equation}
x^{\prime 0}=x^{0}-\tau,
\quad
x^{\prime k}=x^{k}, \quad 
\quad
p^{\prime}_{\mu}=p_{\mu}. 
\end{equation}
The constraint function $\Phi_{1}^{G}$ takes the form 
\begin{eqnarray}
\Phi^{G}_1=x^{\prime 0} 
\end{eqnarray}
and the new Hamiltonian is given by
\begin{equation}
\label{hamiltonian3}
H^{(3)}=-p_0 -\frac{1}{2} ep^{2}-u\left( hp_{0}+\frac{i}{2}
\epsilon_{ijk}p^{i}\eta^{j}\eta^{k}\right)
+\lambda^{1}\Phi^{(1)}_{1}.
\end{equation}
The conservation condition for the constraint $\Phi^G_1$ implies
\begin{equation}
\Phi^{G}_{2}=0 \qquad \textrm{where}\qquad \Phi^{G}_{2}=ep_{0}+ uh+1.
\end{equation}
The consistency condition for this constraint yields $\lambda^1=0$.
One can replace the set of  constraints obtained  by the equivalent set 
$\Phi$,
\begin{eqnarray}
\label{constraints-final}
&& \Phi_{1}=P_{e}, \;\phantom{xxxxi}\Phi_{2}=ep_0+1,\quad \Phi_{3}=P_{u},
\quad \Phi_{4}=u, 
\nonumber \\
&&\Phi_{5}=p^2, \;  \Phi_{6}=x^{\prime 0}, 
\; \Phi_{7}=P_{h}, \;\, \Phi_{8}=hp_0+\frac{i}{2}\epsilon_{ijk}p^{i}\eta^{j}\eta^{k}. 
\end{eqnarray}
This is a set of second-class constraints as a direct calculation proves. 

There are seventeen fundamental variables at this stage and eight constraint equations. Apparently, nine variables should be considered independent and the rest of the varibles can be expressed in terms of the independent ones. However, one of our constraints is quadratic.

\subsection{Dirac brackets}

The nonzero Dirac brackets of the canonical variables with respect to the set of constraints $\Phi$, eq.~(\ref{constraints-final}),   are
\begin{eqnarray}
\label{xpeta}
&&\left\{x^{i},p_{j}\right\}_{D(\Phi)}=\delta_{j}^{i},\qquad
\left\{\eta^{i},\eta^{j}\right\}_{D(\Phi)}=-2i\delta^{ij},
\nonumber\\
&&\left\{x^{i}, p_0\right\}=\frac{p_{i}}{p_{0}}, \qquad
\left\{x^{i},e\right\}_{D(\Phi)}=\frac{p_{k}}{p_{0}\mathbf{p}^{2}}
\nonumber\\
&&\left\{x^{i},h\right\}_{D(\Phi)}=\frac{i}{2p_{0}}
\epsilon_{klm}\left(\delta_{i}^{k}+\frac{p^{k}p_{i}}{\mathbf{p}^{2}}\right)\eta^{l}\eta^{m}, 
\nonumber\\
&&\left\{\eta^{i},h\right\}_{D(\Phi)}=\frac{2}{p_{0}}
\epsilon_{ijk}p^{j}\eta^{k}.
\end{eqnarray}
Only the term $-p_0$ in the Hamiltonian $H^{(3)}$ does not vanish on the constraint surface. Then the Hamiltonian 
\begin{equation}
\label{hamiltonian4}
{H}_{\textrm{phys}}=-p_0
\end{equation}
can be used instead oh $H^{(3)}$. The evolution of the quotient algebra elements is consitently defined by the Hamiltonian equations. However, the Dirac brackets (\ref{xpeta}) do not possess canonical form. This was to be exepcted, because the variables $p^{\mu}$ are not independent: they are related by the constraint equation $\Phi_{5}=0$.  

\subsection{Solutions in the complex algebra}
If the pseudoclassical algebra is just a $\mathbb{C}$-module, only two solutions exist for the equation $\Phi_{5}=0$, namely,
\begin{equation}
\label{energy}
p_{0}=-\zeta\sqrt{\mathbf{p}^{2}}, 
\end{equation}
where $\zeta=\pm 1$ is the energy sign.
Putting $\zeta=1$ one obtains, upon quantization, the quantum mechanics of the spinning massless particles with both signs of helicity. The second solution yields the quantum mechanics of the negative-energy states with both signs of helicity. 
The chirality $h$ is equal to helicity,
\begin{equation}
\label{henergy}
h=
\frac{i}{2\sqrt{\mathbf{p}^{2}}}\zeta \epsilon_{ijk}p^{i}\eta^{j}\eta^{k}.
\end{equation}
and is a conserved quantity. Indeed, its Dirac bracket with the Hamiltonian
\[
H_{\zeta}=\zeta\sqrt{\mathbf{p}^{2}}
\]
vanishes.

\subsection{Quaternionic pseudoclassical mechanics}
Since our aim is to reproduce, by strictly canonical methods, the quantum mechanics of the chiral sectors, we assume that the pseudoclassical algebra $\mathcal{A}$ is a $(\mathbb{B}, \mathbb{B})$-bimodule, where $\mathbb{B}$ is the algebra of the biquaternions. Further, we postulate the following relation between the left and right multiplication 
of the algebra elements by quaternions $\mathbb{H}$. Let $\{1, J^{1},J^{2},J^{3}\}$ be a basis in $\mathbb{H}$, such that  
\begin{equation}
\label{quaternions}
J^{i}J^{j}=-\delta^{ij}+\epsilon_{ijk}J^{k}.
\end{equation}
Then 
\begin{equation}
\label{leftright}
J^{i}\eta^{k}=\left\{
\begin{array}{rll}
-\eta^{k}J^{i}& \textrm{if} & i\not=k\\
\eta^{k}J^{i}& \textrm{if} & i=k
\end{array}
\right.
\end{equation}
The rest of the generating elements of $\mathcal{A}$, as well as the evolution parameter $\tau$, commute with $J^{k}$. These rules are invariant under a change of the basis in $\mathbb{H}$ 
\[
\{1, J^{1},J^{2},J^{3}\}\quad \rightarrow \quad\{1, J^{\prime 1},J^{\prime 2},J^{\prime 3}\},
\] 
if $J^{\prime 1}$ ,$J^{\prime 2}$, $J^{\prime 3}$ satisfy eq.~(\ref{quaternions}), because the quotient algebras arising when the two bases are used, are isomorphic.  

Equation $p^2=0$ possesses new solution in the extended algebra, in addition to the solutions  (\ref{energy}).  These are
\[
p_0=-\chi iJ^{k}p_{k},
\]   
where $J^{k}\in \mathbb{H}$ satisfy (\ref{quaternions}). The constant $\chi=\pm 1$ is the (pseudoclassical counterpart) of helicity (multiplied by two).  
Then 
\[
p_0^{-1}=-\frac{\chi}{\mathbf{p}^2}iJ^{k}p_k.
\]
Because of the  relation postulated (\ref{leftright}) between the left and the right multiplication by quaternions, the constraint equation $\Phi_8=0$ is ambiguous without specifying the order of the factors in its right-hand side. Assuming that $h$ stays to the left of $p_0$, one obtains
\[
h=\frac{i}{2}\epsilon_{ijk}p^{i}\eta^{j}\eta^{k}p_0^{-1}=
\frac{\chi}{2\mathbf{p}^2}\epsilon_{ijk}p^{i}\eta^{j}\eta^{k}(J^lp_l).
\]  
The quantity $h$ is conserved, since its Dirac bracket with the Hamiltonian
\begin{equation}
\label{hamiltonian5}
H={i}\chi J^{k}p_k
\end{equation}
vanishes. 

\section{Quantization}

According to Dirac rules, the operators $\hat{x}^k$, $\hat{p}_k$, and $\hat{\eta}^k$, corresponding to
the independent variables ${x}^k$, ${p}^k$, and $\eta^k$, must satisfy the (anti-)commutation relations  
\begin{eqnarray}
\label{commutators}
\left[\hat{x}^{i},\hat{p}_{j}\right]_{-}&=&i\delta_{j}^{i}, \\
\label{anticommutators}
\left[\hat{\eta}^{i},\hat{\eta}^{j}\right]_{+}&=&2\delta^{ij}.
\end{eqnarray}

\subsection{Positive- and negative-energy sectors}
Let $\mathcal{H}_{xp}$ be a Hilbert space in which an irreducible representation, generated by $\hat{x}^{i}$, $\hat{p}_i$,  of the Heisenberg algebra is given.  Let $\mathcal{H}_{\eta}$ be a unitary space in which  an irreducible representation, generated by $\hat{\eta}^{i}$, of the Clifford algebra is given. Then in $\mathcal{H}=\mathcal{H}_{\eta}\otimes \mathcal{H}_{xp}$ an irreducible representation of the Lie superalgebra is given. 
If the solution (\ref{henergy}) is  chosen, then
\[
\hat{p}_0=-\zeta\sqrt{\hat{\mathbf{p}}^{2}}.
\]
The energy sign operator 
\[
\hat{\zeta}=-\frac{\hat{p}_0}{\sqrt{\hat{\mathbf{p}}^{2}}}
\]
is a number, and, therefore, it commutes with all the canonical operators.
The rest of the dependent operators,  $\hat{h}$  and $\hat{e}$, are expressed in terms of the independent ones,
\begin{equation}
\label{dependent}
\hat{h}=\frac{i{\zeta}}{2\sqrt{\hat{\mathbf{p}}^{2}}}
\epsilon_{ijk}\hat{p}^{i}\hat{\eta}^{j}\hat{\eta}^{k},\quad
\hat{e}=\frac{\zeta}{\sqrt{\mathbf{p}^{2}}},
\end{equation}
according to eqs.~(\ref{constraints-final}), and the Hamiltonian is given by
\begin{equation}
\label{hamiltonian6}
\hat{H}_{{\zeta}}=\zeta\sqrt{\hat{\mathbf{p}}^2}.
\end{equation}
The Schr\"{o}dinger equation reads
\begin{equation}
\label{Schroedinger}
i\frac{\partial}{\partial x^0}\mid \phi(x^{0})\rangle=\zeta\sqrt{\hat{\mathbf{p}}^2}\mid\phi(x^{0})\rangle, 
\end{equation}
where $\mid\phi(x^{0})\rangle$ is a time-dependent vector in $\mathcal{H}$.
An explicit realization is given in  the Hilbert space of the (two-component, square-integrable) spinor-valued functions $\phi(\mathbf{x})$, where one can put  
\begin{equation}
\hat{x}^k=x^k,\qquad \hat p_k=-i\frac{\partial}{\partial {x}^k},
\qquad \hat{\eta^{k}}=\sigma^k,  
\end{equation}
$\sigma^k$ being the Pauli matrices. Taking the direct sum of two copies of this Hilbert space and putting $\zeta=1$ for the first copy and $\zeta=-1$ for the second, one obtains 
\[
\hat{\zeta}=\beta\equiv{\gamma^{0}}, \qquad \hat{\eta^{k}}=\gamma^{5}\gamma^{0}\gamma^{k}.
\]
The Schr\"{o}dinger equation for this system is the Dirac equation in the Foldy-Wouthuisen form,
\begin{equation}
\label{Foldy}
i\frac{\partial}{\partial x^0} \psi(x)=\beta\sqrt{\hat{\mathbf{p}}^2}\psi(x). 
\end{equation}
and the chirality  operator $\hat{h}$ reads 
\[
\hat{h}=\frac{i}{2\sqrt{\hat{\mathbf{p}}^{2}}}\gamma^{0}
\epsilon_{ijk}\hat{p}^{i}\hat{\eta}^{j}\hat{\eta}^{k}=-\frac{1}
{\sqrt{\hat{\mathbf{p}}}^{2}}\gamma^{5}\gamma^{k}\hat{p}_{k}.
\]
It anticommutes with the operators $\hat{\eta}^{k}$ (and commutes with $\hat{x}^{k}$, $\hat{p}_{k}$), while the energy-sign operator $\hat{\zeta}$ commutes with all the canonical operators. 

\subsection{Chiral sectors}

Let $a=a_{0}+a_{i}J^{i}\in \mathbb{B}$ be, where $a_{0}, a_{i}\in \mathbb{C}$. In biquaternionic algebra we can define three conjugations, namely

\[
a^{\ast}=a_{0}^{\ast}+a_{i}^{\ast}J^{i},
\]

\[
\bar{a}=a_{0}-a_{i}J^{i},
\]

\[
a^{\dagger}=\bar{a}^{\ast}=a_{0}^{\ast}-a_{i}^{\ast}J^{i}.
\]
These are complex, quaternionic and hermitean conjugations, respectively. With these operations, we can define biquaternionic scalar product as

\[
(a,b)=Re(a^{\dagger}b).
\]
Following \cite{mori83} we have
\[
Re(z)=\frac12(z+\bar{z}),
\]
and also $(\overline{zw})=\bar{w}\bar{z}$, $\overline{\bar{w}}=w$. Then we can rewrite the scalar product as
\be
(a,b)=\frac12(\bar{a}^{\ast}b+\bar{b}a^{\ast}).
\ee

The operators $\hat{\eta^{k}}$, acting in $\mathbb{B}$ and defined by 
\[
\hat{\eta^{k}}=iJ^{k},  
\]
satisfy the anticommutation relations (\ref{anticommutators}) and posses the same commutation properties with $J^{k}$ as the pseudoclassical variables $\eta^{k}$ do. The representation of the Clifford algebra in $\mathbb{B}$, generated by these operators is reducible and decomposes into two irreducible representations. The subspace $\mathbb{B}_{+}$ , spanned on the vectors 
\[
\mid 1\rangle = \frac{1}{\sqrt{2}}\left(1+iJ^{3}\right), \qquad 
\mid 2\rangle = \frac{1}{\sqrt{2}}\left(J^{2}+iJ^{1}\right).
\]
is one of the invariant subspaces. The representation  of the Clifford algebra in $\mathbb{B}_{+}$, generated by $\hat{\eta}^{k}$ is irreducible. Let an irreducible representation, generated by $x^{k}$ and $p_{k}$,  of the Heisenberg algebra be given in some complex Hilbert space $\mathcal{H}_{xp}$. An irreducible representation of the canonical (anti-) commutation relations is defined in the tensor product  $\mathcal{H}=\mathbb{B}_{+}\otimes\mathcal{H}_{xp}$ in a natural way.  The Hamiltonian is given by 
$\hat{H}={i}\chi J^{k}\hat{p}_k$, according to eq.~(\ref{hamiltonian5}).

Let ${\mid \mathbf{x}\rangle}$ be the basis of the coordinate eigenvectors in $\mathcal{H}_{xp}$,
\[
\hat{x}^{k}\mid \mathbf{x}\rangle =x^{k}\mid \mathbf{x}\rangle, 
\quad \langle \mathbf{x}\mid \hat {p}_{k}\mid \phi\rangle=-i\frac{\partial}{\partial x^{k}}\langle \mathbf{x}\mid \phi\rangle, \quad 
\langle \mathbf{x}\mid \mathbf{x}^{\prime}\rangle=
\delta^{3}(\mathbf{x}-\mathbf{x}^{\prime}).
\] 
A basis in $\mathcal{H}$ is given by the vectors 
$\mid \alpha, \; \mathbf{x}\rangle=\mid \alpha\rangle\otimes \mid\mathbf{x}\rangle$ ($\alpha=1,2$).
A state vector $\mid\psi \rangle$ in $\mathcal{H}$ is represented in this basis by a two-component wave function $\psi$,
\[
\psi(\mathbf{x})=\left(
\begin{array}{c}
\langle 1, \mathbf{x}\mid \psi\rangle \\
\langle 2, \mathbf{x}\mid \psi\rangle 
\end{array}
\right)
\] 
Since 
\[
\left(
\begin{array}{c}
\langle 1, \mathbf{x}\mid \hat{\eta}^{k}\mid \psi\rangle \\
\langle 2, \mathbf{x}\mid \hat{\eta}^{k}\mid \psi\rangle 
\end{array}
\right)=\sigma^{k}\psi(\mathbf{x}),
\]
the time dependent wave function satisfies the Weyl equation
\[
i\frac{\partial}{\partial x^{0}}\psi(\mathbf{x}) =-i\chi \sigma^{k}\frac{\partial}{\partial x^{k}}\psi(\mathbf{x}).
\]

\section*{Acknowledgements}

The authors thank A. Das, D. M. Gitman and R.~Fresneda for discussions.

M.N.B thanks Departamento de Física da Universidade Federal da Paraíba and S.I.Z thanks Conselho Nacional de Desenvolvimento Ci\^ent\'i\-fi\-co e Tecnol\'ogico (CNPq) of Brazil for financial support.

\end{document}